\documentclass[reprint, amsmath, amssymb, aps, pra]{revtex4-2}

\usepackage{subfigure}
\usepackage{amsmath}
\usepackage{enumitem}
\usepackage{graphicx}
\usepackage{dcolumn}
\usepackage{bm}
\usepackage[colorlinks=true,allcolors=blue]{hyperref}
\usepackage{booktabs} 
\usepackage{makecell}

\usepackage{xcolor}

\newcommand{\rev}[1]{#1}

\begin{document}

\title{Topological enhancement of a $\mathcal{PT}$-symmetric Su-Schrieffer-Heeger quantum battery}

\author{A-Long Zhou}

\author{Ya-Wen Xiao}

\author{Nuo Xu}

\author{Li-Li Gao}

\author{Long-Jie Li}

\author{Hang Zhou}

\author{Zi-Min Li}
\email{zimin.li@csu.edu.cn}

\author{Chuan-Cun Shu}
\email{cc.shu@csu.edu.cn}

\affiliation{Institute of Quantum Physics, Hunan Key Laboratory of Nanophotonics and Devices, Hunan Key Laboratory of Super-Microstructure and Ultrafast Process, School of Physics, Central South University, Changsha 410083, China}

\date{\today}

\begin{abstract}
We investigate a non-Hermitian quantum battery based on the Su-Schrieffer-Heeger (SSH) lattice, charged through a parity-time (\(\mathcal{PT}\))-symmetric protocol that alternates gain and loss between the two sublattices. 
The interplay between lattice topology and non-Hermiticity gives rise to both bulk and edge exceptional points (EPs), which govern the charging dynamics.
In the topological regime, an edge-state EP appears at a smaller gain-loss strength than the bulk thresholds and gives rise to an additional edge-broken regime absent in the trivial configuration.
This topology-specific spectral structure is reflected in the charging dynamics, where the topological phase exhibits more favorable transient and long-time performance in the representative non-Hermitian regimes considered here.
We further examine the corresponding Lindblad dynamics, identifying the non-Hermitian model as the conditional no-jump description of the same gain-loss processes.
The Lindblad results show that the topological advantage remains visible at the level of stored energy, extractable work, and extractable fraction under unconditional open-system evolution.
These findings demonstrate that topology constitutes a genuine physical resource for enhancing the performance of quantum batteries.
\end{abstract}

\maketitle

\section{Introduction}

Quantum batteries (QBs) are quantum systems that store and deliver energy by exploiting inherently quantum resources such as coherence, entanglement and collective interactions \cite{Campaioli2024, Pokhrel2025, simon_correlations_2025, gyhm_beneficial_2024, mayo_collective_2022, li_switchable_2025, shi_entanglement_2022, alicki_entanglement_2013, Campaioli_Enhancing_2018}.
They have become an active research topic at the interface of quantum information and thermodynamics.
Various QB models have demonstrated advantages in charging power, efficiency, and robustness over classical counterparts.
For example, collective charging schemes in Dicke or Tavis–Cummings models show how cooperative quantum effects can increase power beyond the linear scaling of independent units \cite{Ferraro2018, seidov_quantum_2024, fusco_work_2016, dou_extended_2022, crescente_ultrafast_2020, andolina_quantum_2019, ghosh_enhancement_2020, Su_Quantum_2024}.
Nevertheless, identifying additional physical mechanisms that can enhance charging speed or stability under realistic conditions remains a central challenge in the development of practical quantum batteries \cite{Campaioli_Enhancing_2018, Dou_Quantum_2023}.

Beyond quantum correlations, topological phases of matter offer an alternative approach to controlling energy storage and transfer.
Their global invariants protect edge states and yield robust phase transitions, suggesting that topological features could enhance the stability and efficiency of quantum-battery performance \cite{Hasan2010,Asboth2016}.
Among the simplest one-dimensional realizations is the Su–Schrieffer–Heeger (SSH) model, a dimerized lattice with alternating couplings that serves as a minimal platform for exploring topology in charging dynamics \cite{Su1979}.

Recent studies have begun to explore the SSH model in the context of energy storage. 
For instance, Ref.~\cite{Zhao2022a} considered a spin-chain QB with SSH-type couplings and examined how a quantum phase transition influences the charging performance, but without analyzing the explicit role of topology.
More recently, Ref.~\cite{Lu2025a} introduced the concept of topological quantum batteries and showed that, in open systems, non-trivial topological phases support bound states that suppress dissipation and allow the system to maintain stored energy over long times.
Building on these studies, we examine the SSH lattice itself as a quantum battery, emphasizing how topological phases modify charging efficiency and stability.

While many QB proposals assume isolated unitary dynamics, in practice no quantum system is completely free from environmental effects. 
Dissipation and decoherence play an inevitable role and must therefore be incorporated into realistic models of QBs \cite{Guo2025, Pokhrel2025, carrasco_collective_2022, zhang_dissipative_2025, zhao_quantum_2021, ghosh_fast_2021, quach_using_2020, hovhannisyan_charging_2020}. 
A standard theoretical framework for describing such open-system effects is the Lindblad master equation, which enables the treatment of non-unitary evolution in a controlled manner \cite{Manzano_2020, wilma2018visualizing}. 
In certain regimes, however, these open dynamics can be effectively captured by non-Hermitian Hamiltonians, leading to simplified yet insightful descriptions \cite{ElGanainy2018, Ashida2020,Li2024a}.
Recent studies have shown that non-Hermitian dynamics can themselves enhance the performance of quantum batteries \cite{Konar2024}. 
By engineering gain and loss, one can achieve faster charging or more robust energy storage compared with purely Hermitian setups. 
\rev{A related open-system perspective was considered by Barra \textit{et al.}~\cite{Barra2022}, who studied a cyclic battery--charger device operating close to a quantum phase transition in a topologically trivial setting.}
\rev{They showed that the extracted work and thermodynamic efficiency can be enhanced near criticality.}
\rev{These results further illustrate that quantum-battery performance can be strongly influenced by the spectral and dynamical structure of the underlying open-system setting.}
A particularly intriguing case is that of parity-time (\(\mathcal{PT}\)) symmetric Hamiltonians, which balance gain and loss in a way that preserves real spectra in certain regimes \cite{Bender_2005, Dogra_2021, klett_relation_2017, zhu_pt_2014}. 
The role of \(\mathcal{PT}\) symmetry in quantum batteries remains largely unexplored. 
It is therefore natural to ask how \(\mathcal{PT}\)-symmetric extensions of the SSH model may compete or cooperate with topological effects in determining charging performance.

From an experimental perspective, the key ingredients of non-Hermitian and topological lattice models have already been realized in several platforms.
Photonic lattices and coupled-resonator arrays can implement SSH-type band structures with tunable gain and loss, enabling direct access to $\mathcal{PT}$ symmetry and exceptional-point physics~\cite{ElGanainy2018,Oezdemir2019,Zeuner_Observation_2015}.
In addition, cavity- and circuit-QED architectures provide versatile settings in which driven light--matter interactions realize generalized quantum Rabi and lattice models with effective non-Hermitian and topological features~\cite{Braak_2011,Li2021a,Li2021GAA,Xie_2017,Li_2015,Xie_2014,Lu2023,Fan2023}.
These developments indicate that the $\mathcal{PT}$-symmetric topological mechanisms explored here could be transplanted into cavity-based quantum-battery charging schemes, where an SSH-type battery is coupled to structured photonic modes in a controllable way.

\rev{In this work, we investigate a \(\mathcal{PT}\)-symmetric SSH quantum battery that integrates topological and non-Hermitian effects.}
\rev{The interplay between topology and \(\mathcal{PT}\) symmetry gives rise to both bulk and edge exceptional points (EPs), leading to a topology-dependent \(\mathcal{PT}\)-breaking structure and distinct charging dynamics in the topological and trivial phases.}
\rev{In particular, the topological phase exhibits an additional edge-broken regime that is absent in the trivial case, and this topology-specific spectral structure is reflected in its charging behavior.}
\rev{We further examine the corresponding Lindblad dynamics and identify the non-Hermitian model as the conditional no-jump description of the same gain-loss processes.}
\rev{The Lindblad results show that the topological advantage remains visible at the level of stored energy, extractable work, and extractable fraction under unconditional open-system evolution.}
\rev{These results highlight the cooperative role of topology and non-Hermiticity in shaping quantum-battery performance.}

\section{Su--Schrieffer--Heeger (SSH) quantum battery}
\label{sec:sshqb}

\begin{figure}[t]
  \centering
  \includegraphics[width=.95\linewidth]{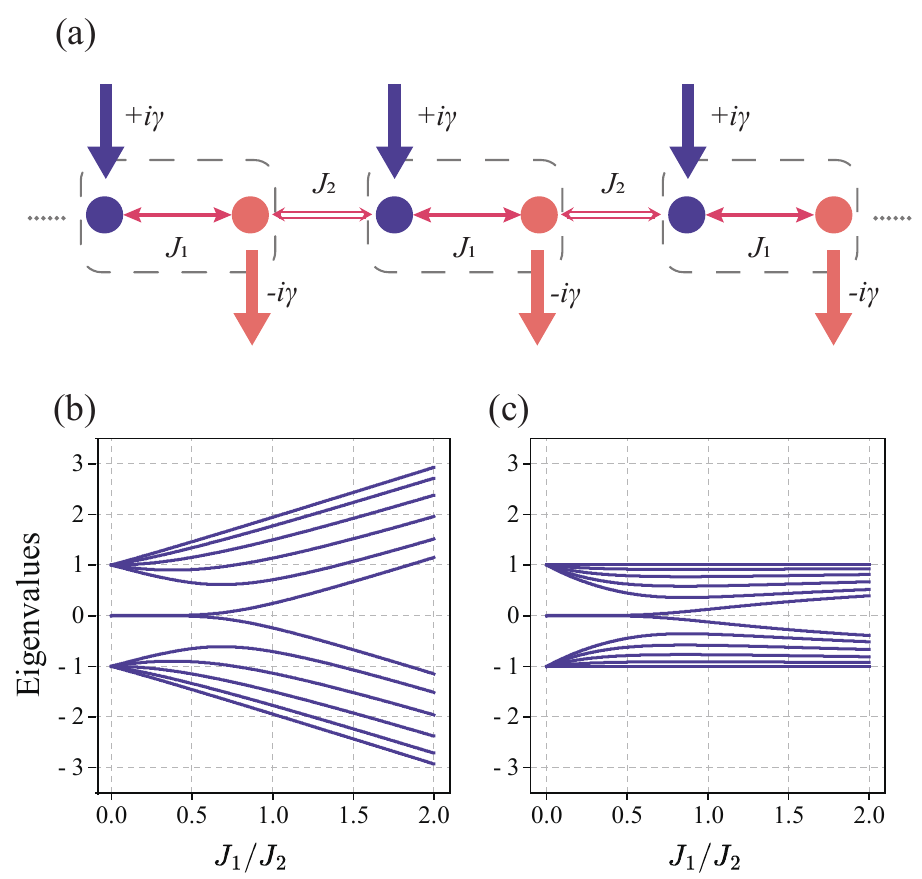}
  \caption{
    (a) SSH lattice with alternating couplings $J_{1}$ and $J_{2}$, serving as the basis for the $\mathcal{PT}$-symmetric SSH model with balanced gain and loss $\gamma$.
    \rev{(b) Single-particle spectrum of the Hermitian SSH model under open boundary conditions.}
    (c) Normalized energy spectrum, used to eliminate the trivial dependence on the overall energy scale and to enable consistent evaluation of charging performance.
}
  \label{fig:ssh_setup}
\end{figure}

\subsection{SSH model and chiral symmetry}
\label{sec:SSHmodel}

We consider a one-dimensional dimerized chain of $2N$ sites arranged in $N$ unit cells, each containing an $A$ and a $B$ sublattice site.
\rev{
For an SSH chain with open boundary conditions and \(N\) unit cells, the Hermitian Hamiltonian is
}
\begin{equation}
    \rev{
    H_{\mathrm{SSH}}
    =
    \sum_{n=1}^{N}
    \Big(
    J_{1}a_{n}^{\dagger}b_{n}
    + H.c.
    \Big)
    +
    \sum_{n=1}^{N-1}
    \Big(
    J_{2}b_{n}^{\dagger}a_{n+1}
    + H.c.
    \Big),
    }
    \label{eq:H_SSH}
\end{equation}
where $a_{n}$ and $b_{n}$ annihilate excitations on sublattices $A$ and $B$ of cell~$n$, respectively, and $J_{1(2)}\!\ge0$ denote the intra-(inter-)cell couplings.

The model possesses chiral (sublattice) symmetry generated by
\begin{equation}
    \Gamma=\sum_{n}(a_{n}^{\dagger}a_{n}-b_{n}^{\dagger}b_{n}),
    \label{eq:Gamma_def}
\end{equation}
which anticommutes with $H_{\mathrm{SSH}}$, $\{\Gamma,H_{\mathrm{SSH}}\}=0$ and therefore ensures that the spectrum is symmetric about zero energy ($E\!\leftrightarrow\!-E$), see Appendix~\ref{app:chiral} for further discussion.

Under periodic boundary conditions and Bloch transformation $a_{k}^{\dagger}=N^{-1/2}\!\sum_{n}e^{ikn}a_{n}^{\dagger}$ and $b_{k}^{\dagger}=N^{-1/2}\!\sum_{n}e^{ikn}b_{n}^{\dagger}$, the Bloch Hamiltonian becomes
\begin{equation}
    h(k)=
    \begin{pmatrix}
        0 & J_{1}+J_{2}e^{-ik}\\
        J_{1}+J_{2}e^{ik} & 0
    \end{pmatrix},
\end{equation}
whose bulk spectrum
\begin{equation}
        E_{\mathrm{SSH}}(k)=\pm\sqrt{J_{1}^{2}+J_{2}^{2}+2J_{1}J_{2}\cos k}
\end{equation}
closes its gap at $J_{1}=J_{2}$, which marks the topological transition.
For open boundaries, the chain hosts zero-energy edge states when $J_{1}<J_{2}$, corresponding to a dimerization centered on the weaker bonds, while it becomes topologically trivial for $J_{1}>J_{2}$.

The SSH lattice configuration is sketched in Fig.~\ref{fig:ssh_setup}(a), which also forms the basis for the $\mathcal{PT}$-symmetric extension introduced in Sec.~\ref{sec:PT-spectrum}. 
\rev{Figure~\ref{fig:ssh_setup}(b) shows the corresponding single-particle spectrum of the Hermitian SSH model under open boundary conditions, while the normalized spectrum in Fig.~\ref{fig:ssh_setup}(c) will be discussed later in Sec.~\ref{sec:dynamics}.}
All the figures are generated with $N=6$.

\subsection{$\mathcal{PT}$-symmetric SSH model}
\label{sec:PT-spectrum}

We now consider a charging scheme with balanced gain and loss on the two sublattices, effectively described by a $\mathcal{PT}$-symmetric extension of the SSH lattice introduced above. 
The corresponding non-Hermitian Hamiltonian is 
\begin{equation}
H_{\mathcal{PT}}
  = H_{\mathrm{SSH}}
    + i\gamma\sum_{n=1}^{N}(a_{n}^{\dagger}a_{n}-b_{n}^{\dagger}b_{n})
  \equiv H_{\mathrm{SSH}}+i\gamma\,\Gamma,
\label{eq:H_PT}
\end{equation}
where $\gamma > 0$ quantifies the strength of the gain–loss coupling \cite{Tzortzakakis2022}. 
The Hamiltonian commutes with the combined $\mathcal{PT}$ operator, where $\mathcal{P}$ exchanges the sublattices and $\mathcal{T}$ denotes complex conjugation. 
This ensures a real spectrum in the unbroken $\mathcal{PT}$-symmetric regime. 
The effective non-Hermitian Hamiltonian can be derived from the Lindblad master equation by neglecting the quantum-jump terms, as detailed in Section \ref{sec:lindblad_validation}. 

Under periodic boundary conditions, the bulk dispersion in momentum space reads
\begin{equation}
E_{\mathcal{PT}}(k) 
  = \pm \sqrt{ J_{1}^{2} + J_{2}^{2} + 2J_{1}J_{2}\cos k - \gamma^{2} } .
\end{equation}
\rev{The corresponding bulk EPs occur at $\gamma = |J_{1}\!\pm\! J_{2}|$, where pairs of eigenvalues coalesce and the bulk spectrum changes from real to complex, marking the spontaneous breaking of $\mathcal{PT}$ symmetry.}
\rev{These bulk thresholds separate three bulk $\mathcal{PT}$-symmetric regimes in the thermodynamic limit: unbroken ($\gamma < |J_{1}-J_{2}|$), partially broken ($|J_{1}-J_{2}| < \gamma < |J_{1}+J_{2}|$), and fully broken ($\gamma > |J_{1}+J_{2}|$).}
\rev{For the finite open chain in the topological phase, the hybridized edge-state pair experiences an additional EP at}
\begin{equation}
    \gamma_{e}=J_1\frac{1-(J_1/J_2)^2}{1-(J_1/J_2)^{2N}}(J_{1}/J_{2})^{N-1}, 
    \label{eq:gamma_e}
\end{equation}
\rev{which decreases exponentially with the system size $N$ and vanishes in the trivial phase~\cite{Tzortzakakis2022}.}
\rev{This edge EP originates from the coalescence of the two midgap edge modes and is a characteristic feature of the topological regime.}
\rev{Here, however, $\gamma_e$ and $\gamma=|J_{1}\!\pm\!J_{2}|$ play different roles: $\gamma_e$ is the finite-size edge-EP estimate from the effective two-state description of the open chain, whereas $\gamma=|J_{1}\!\pm\!J_{2}|$ are the bulk threshold scales inherited from the periodic thermodynamic-limit spectrum.}

\begin{figure}[t]
  \centering
  \includegraphics[width=.9\linewidth]{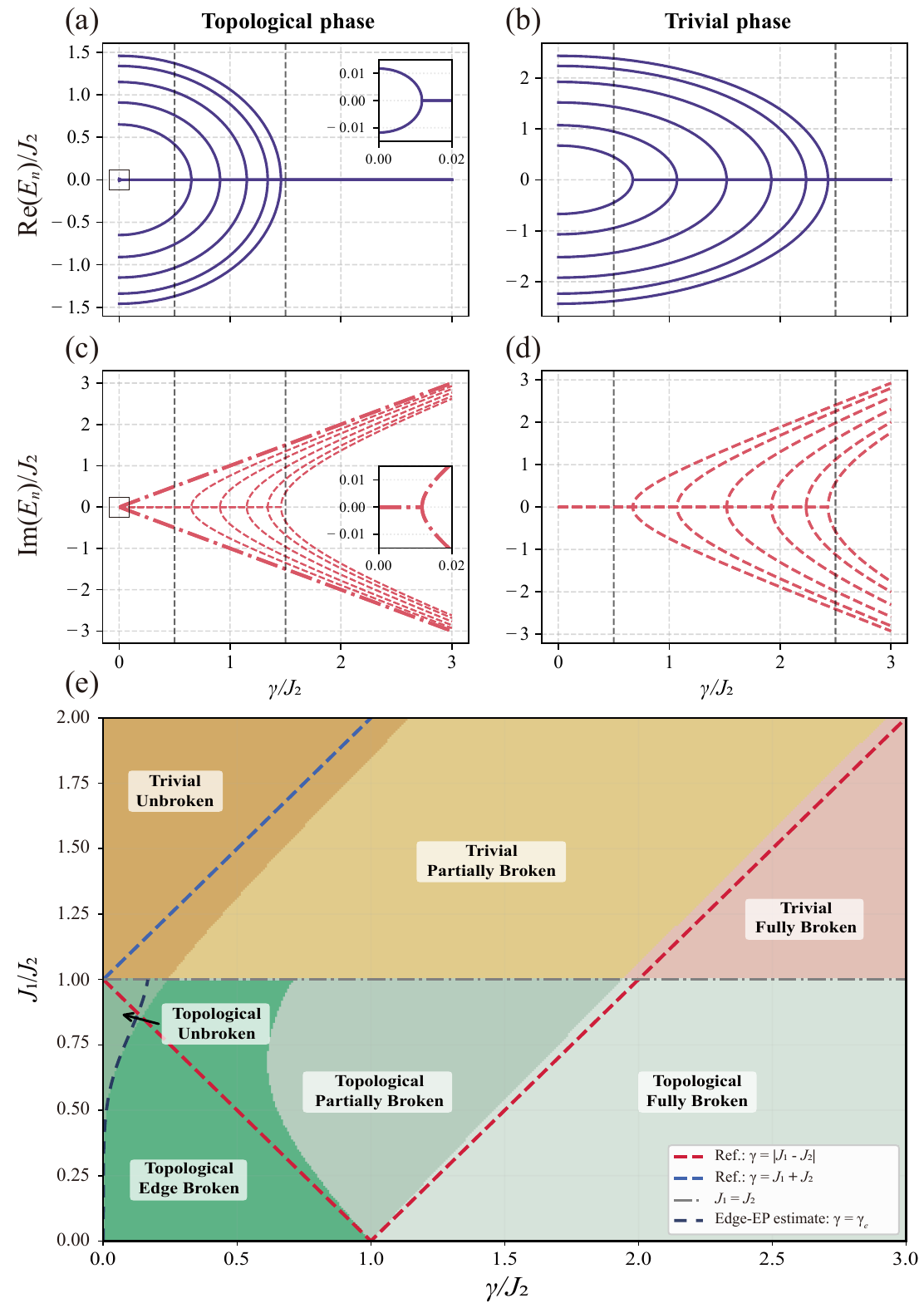}
    \caption{
    Spectral structure and regime diagram of the $\mathcal{PT}$-symmetric SSH model.
    (a,c) Real and imaginary parts of the eigenvalues in the topological phase ($J_{1}/J_{2}=0.5$).
    (b,d) Corresponding spectra in the trivial phase ($J_{1}/J_{2}=1.5$).
    \rev{(e) Finite-chain regime diagram in the $\gamma$--$J_{1}$ plane for $N=6$ under open boundary conditions, where the different finite-chain regimes are represented by the colored regions.
    The dashed red and blue lines indicate the bulk threshold scales $\gamma=|J_{1}-J_{2}|$ and $\gamma=J_{1}+J_{2}$ from the thermodynamic-limit spectrum, respectively.
    The dashed navy line denotes the edge-EP estimate $\gamma=\gamma_{e}$ from the effective two-state description.}
    }
  \label{fig:PTSSH_spectrum_phase}
\end{figure}

Figures~\ref{fig:PTSSH_spectrum_phase}(a)-\ref{fig:PTSSH_spectrum_phase}(d) show the real and imaginary parts of the spectrum as functions of gain-loss strength $\gamma$ for topological ($J_{1}/J_{2}=0.5$) and trivial ($J_{1}/J_{2}=1.5$) phases.
For small $\gamma$, all eigenvalues are real, corresponding to the unbroken $\mathcal{PT}$ phase. 
As $\gamma$ increases, pairs of eigenvalues coalesce and split into complex-conjugate partners at the EPs, signaling the spontaneous breaking of $\mathcal{PT}$ symmetry. 
Beyond each EP, one hybridized mode amplifies while its partner decays, reflecting the non-unitary nature of the dynamics.

\rev{In the topological configuration (\(J_{1}<J_{2}\)), the additional edge EP given by Eq.~(\ref{eq:gamma_e}) is associated with the midgap edge-state pair, and the corresponding edge-mode branch in the imaginary spectrum is highlighted by the thick dot-dashed line in Fig.~\ref{fig:PTSSH_spectrum_phase}(c).}
\rev{For the finite open chain, the regime boundaries in Fig.~\ref{fig:PTSSH_spectrum_phase}(e) are identified directly from the exact spectrum rather than imposed from the thermodynamic-limit bulk thresholds.}
\rev{In the topological phase, the first onset of $\mathcal{PT}$ breaking is associated with the edge-state pair and defines the edge-broken regime.}
\rev{At larger $\gamma$, bulk-like modes also become complex, which marks the onset of the partially broken regime, while the fully broken regime is reached when all eigenvalue pairs have become complex.}
\rev{In the trivial phase, where no edge-state pair exists, the partially broken and fully broken regimes are defined analogously from the onset of bulk $\mathcal{PT}$ breaking and from the point where all eigenvalue pairs have become complex.}
\rev{Accordingly, Fig.~\ref{fig:PTSSH_spectrum_phase}(e) presents an exact finite-chain regime diagram for $N=6$. The different finite-chain regimes are indicated by the colored regions, whereas the dashed red, blue, and navy lines serve only as bulk and edge reference scales rather than regime boundaries.}
\rev{The small mismatch between the exact finite-chain boundaries and the dashed $\gamma_e$ line near $J_{1}/J_{2}\to 1$ reflects the limited accuracy of the effective edge-state description in the regime where edge and bulk states are less well separated.}


\subsection{Non-Hermitian dynamics and performance metrics}
\label{sec:dynamics}

The non-Hermitian Hamiltonian $H_{\mathcal{PT}}$ acts as the charger driving the SSH battery.
Within the effective-Hamiltonian description, the normalized density matrix evolves as \cite{Konar2024}
\begin{equation}
    \rho(t)= \frac{e^{-iH_{\mathcal{PT}}t}\rho(0)e^{iH_{\mathcal{PT}}^{\dagger}t}}
        {\mathrm{Tr}[\,e^{-iH_{\mathcal{PT}}t}\rho(0)e^{iH_{\mathcal{PT}}^{\dagger}t}\,]},
        \label{eq:normalized_rho}
\end{equation}
where the initial density matrix $\rho(0)$ is taken as the ground-state projector of $H_{\mathrm{SSH}}$.

Conventionally, the charging energy, average power, and extractable work (ergotropy) are the key quantities used to characterize quantum-battery performance.
In our system, however, since the dynamics is oscillatory and can become non-stationary in the $\mathcal{PT}$-broken regime, time-averaged power is not a meaningful figure of merit. 
Moreover, for a battery initialized in the ground state and with work extraction restricted to unitaries, the ergotropy $\mathcal{W}(t)$ effectively coincides with $\Delta E(t)$, as shown in Appendix~\ref{app:ergotropy}. 
We therefore focus on the stored energy $\Delta E(t)$, defined by
\begin{align}
    \Delta E(t)=E_{B}(t)-E_{B}(0), \quad E_{B}(t)&=\mathrm{Tr}[H_{B}\rho(t)].
\end{align}

To enable consistent comparison across parameters, we remove trivial spectral dilation by normalizing the battery Hamiltonian \cite{Konar2024, ghosh_fast_2021, Konar_Quantum_2022, Ghosh_Dimensional_2022},
\begin{equation}
H_{B}^{(\mathrm{norm})}
   =\frac{2H_{B}-(E_{\max}+E_{\min})\mathbb{I}}
          {E_{\max}-E_{\min}},
\end{equation}
which confines the spectrum to the interval $[-1,1]$.
This rescaling ensures that performance enhancements originate solely from intrinsic topological effects rather than from artificial spectral scaling.


\section{Charging performance analysis}

\subsection{Charging dynamics in different regimes}
\label{sec:charging_dynamics}

\begin{figure}[t]
    \centering
    \includegraphics[width=.9\linewidth]{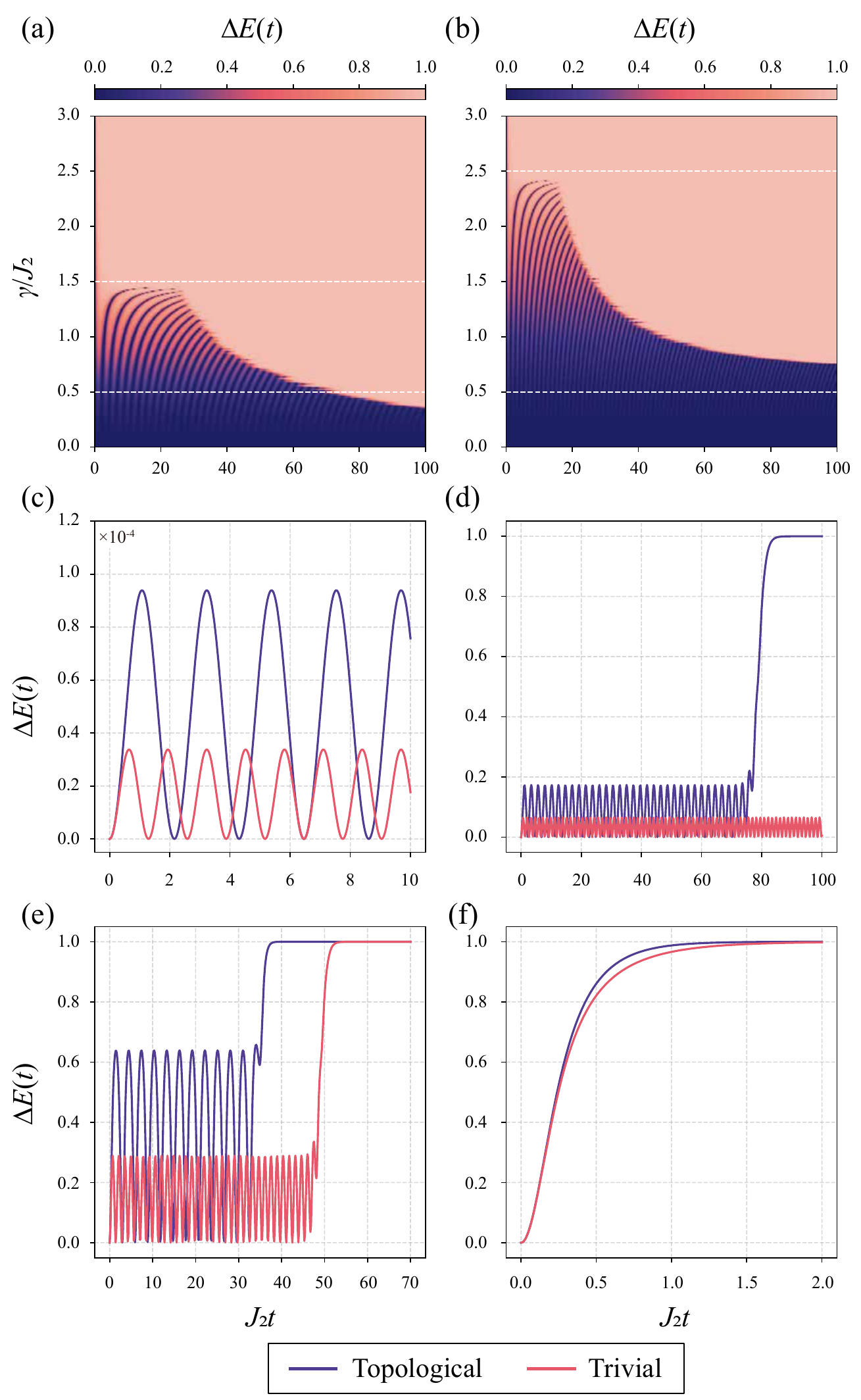}
    \caption{
    Time evolution of the stored energy $\Delta E(t)$ under different non-Hermitian parameters $\gamma$ and topological configurations.
    (a,b) Global maps of $\Delta E(t)$ for the topological ($J_{1}/J_{2}=0.5$) and trivial ($J_{1}/J_{2}=1.5$) phases, respectively.
    (c)--(f) Representative traces of $\Delta E(t)$ at $\gamma/J_{2}=0.01$, $0.45$, $1.0$, and $2.8$, corresponding to the unbroken, edge-state broken, partially broken, and fully broken $\mathcal{PT}$ regimes.
    Purple and red curves denote the topological and trivial phases, respectively.
    \rev{The topological phase generally exhibits faster energy growth and shorter saturation times than the trivial phase in the representative non-Hermitian regimes considered here.}
    }
    \label{fig:charging_dynamics}
\end{figure}

Having established the dynamical model and the performance metric, we now turn to the dynamical behavior of the quantum battery, focusing on the time evolution of the stored energy $\Delta E(t)$ under different non-Hermitian parameters $\gamma$ and topological configurations.

Figures~\ref{fig:charging_dynamics}(a) and \ref{fig:charging_dynamics}(b) show the global evolution of $\Delta E(t)$ for the topological ($J_{1}/J_{2}=0.5$) and trivial ($J_{1}/J_{2}=1.5$) phases, respectively.
The color scale represents the normalized energy $\Delta E(t)$ as a function of both non-Hermitian parameter $\gamma$ and time.
Three distinct dynamical regimes emerge, corresponding to the unbroken, partially broken, and fully broken $\mathcal{PT}$ phases identified in Fig.~\ref{fig:PTSSH_spectrum_phase}.
In the unbroken phase, all eigenvalues remain real and the system exhibits persistent oscillations in $\Delta E(t)$, forming long-lived ripples in the time domain.
In the partially broken phase, the system first displays oscillations and then saturates, leading to a mixed region characterized by a transient oscillatory pattern followed by a steady plateau.
In the fully broken phase, all eigenvalues are complex and $\Delta E(t)$ increases monotonically to its saturation value without oscillations.

\rev{A comparison of Figs.~\ref{fig:charging_dynamics}(a) and~\ref{fig:charging_dynamics}(b), together with the regime analysis in Fig.~\ref{fig:PTSSH_spectrum_phase}(e), shows that the topological phase exhibits a qualitatively different PT-breaking sequence from the trivial phase due to the presence of edge states.}
\rev{Instead of entering directly into the usual bulk partially broken regime, it first passes through an intermediate edge-broken regime in which only the edge-state pair acquires nonzero imaginary parts while the bulk modes remain real.}
\rev{This regime is specific to the topological case and provides an additional topology-dependent distinction from the trivial phase beyond the usual bulk \(\mathcal{PT}\)-breaking classification.}
\rev{As will be seen below, this topology-specific spectral structure already leads to distinct charging behavior in the edge-broken regime, where the bulk modes are still \(\mathcal{PT}\)-unbroken.}

Figures~\ref{fig:charging_dynamics}(c)--\ref{fig:charging_dynamics}(f) illustrate representative time traces of $\Delta E(t)$ for four specific $\gamma$ values, corresponding to the unbroken, edge-broken, partially broken, and fully broken $\mathcal{PT}$ regimes. 

At $\gamma/J_{2}=0.01$ [Fig.~\ref{fig:charging_dynamics}(c)], both phases lie in the unbroken regime, where all eigenvalues remain real and the dynamics is purely oscillatory. 
The stored energy exhibits sustained periodic oscillations with small amplitude, reflecting coherent population exchange between the lattice sites. 
In this limit, the charging behavior of the topological and trivial batteries is nearly identical, and no advantage is observed.

When $\gamma/J_2$ increases to $0.45$ [Fig.~\ref{fig:charging_dynamics}(d)], the trivial phase remains unbroken, whereas the topological phase undergoes edge-state $\mathcal{PT}$ breaking. 
The non-Hermitian growth of the broken edge mode amplifies its population, rapidly transferring energy into the battery sublattice. 
Consequently, the topological battery shows a sharp rise in $\Delta E(t)$ followed by a slow saturation toward a high steady value, while the trivial system continues to oscillate around a much lower mean energy. 
This edge-driven amplification marks the regime where the topological advantage is most pronounced.

Upon further increase to $\gamma/J_{2}=1.0$ [Fig.~\ref{fig:charging_dynamics}(e)], both phases enter the bulk partially broken regime, in which real and complex eigenmodes coexist. 
The early-time oscillations originate from the remaining real spectrum, while the long-time approach to saturation is governed by the dominance of complex-conjugate pairs leading to non-unitary amplification and decay. 
Although the overall charging efficiency begins to level off, the topological configuration still achieves faster convergence.

In the fully broken regime ($\gamma/J_{2}=2.8$) [Fig.~\ref{fig:charging_dynamics}(f)], all eigenvalues become complex and the charging process becomes monotonic. 
The stored energy $\Delta E(t)$ increases steadily toward its maximum attainable value. 
Even under such strong non-Hermiticity, the topological battery retains a slight speed advantage, completing the charge more rapidly and stabilizing earlier than the trivial counterpart.

\rev{Overall, the results show that the topological configuration tends to exhibit more favorable charging behavior than the trivial one across the non-Hermitian regimes considered here.}
\rev{A key difference is the additional edge-broken regime unique to the topological phase, which provides a topology-specific spectral structure reflected in the corresponding charging dynamics.}

\subsection{Quantitative measures of charging performance}
\label{sec:quantitative_measures}

To confirm whether the dynamical behaviors identified above persist across the full parameter space, we introduce two performance indicators extracted directly from the time evolution of $\Delta E(t)$.
These indicators characterize complementary aspects of short- and long-time charging behavior and allow us to evaluate the robustness of the topological advantage under varying non-Hermitian strength $\gamma$ and coupling ratio $J_{1}/J_{2}$.

The first quantity is the amplitude of the first energy peak, denoted $\Delta E_{\mathrm{peak}}^{(1)}$.
This peak arises from the coherent population transfer that dominates the early-time oscillatory regime and therefore measures the battery’s initial responsiveness to the non-Hermitian drive.
A larger value of $\Delta E_{\mathrm{peak}}^{(1)}$ indicates more efficient short-time energy absorption and provides a meaningful comparison even in regions where long-time saturation does not occur.

The second quantity is the saturation time $t_{0.95}$, defined as the earliest time at which the stored energy reaches $95\%$ of its long-time asymptotic value.
This timescale captures how quickly non-unitary amplification and dissipation drive the system toward its steady charging plateau, serving as a natural measure of long-time charging speed.

\begin{figure}[t]
    \centering
    \includegraphics[width=\linewidth]{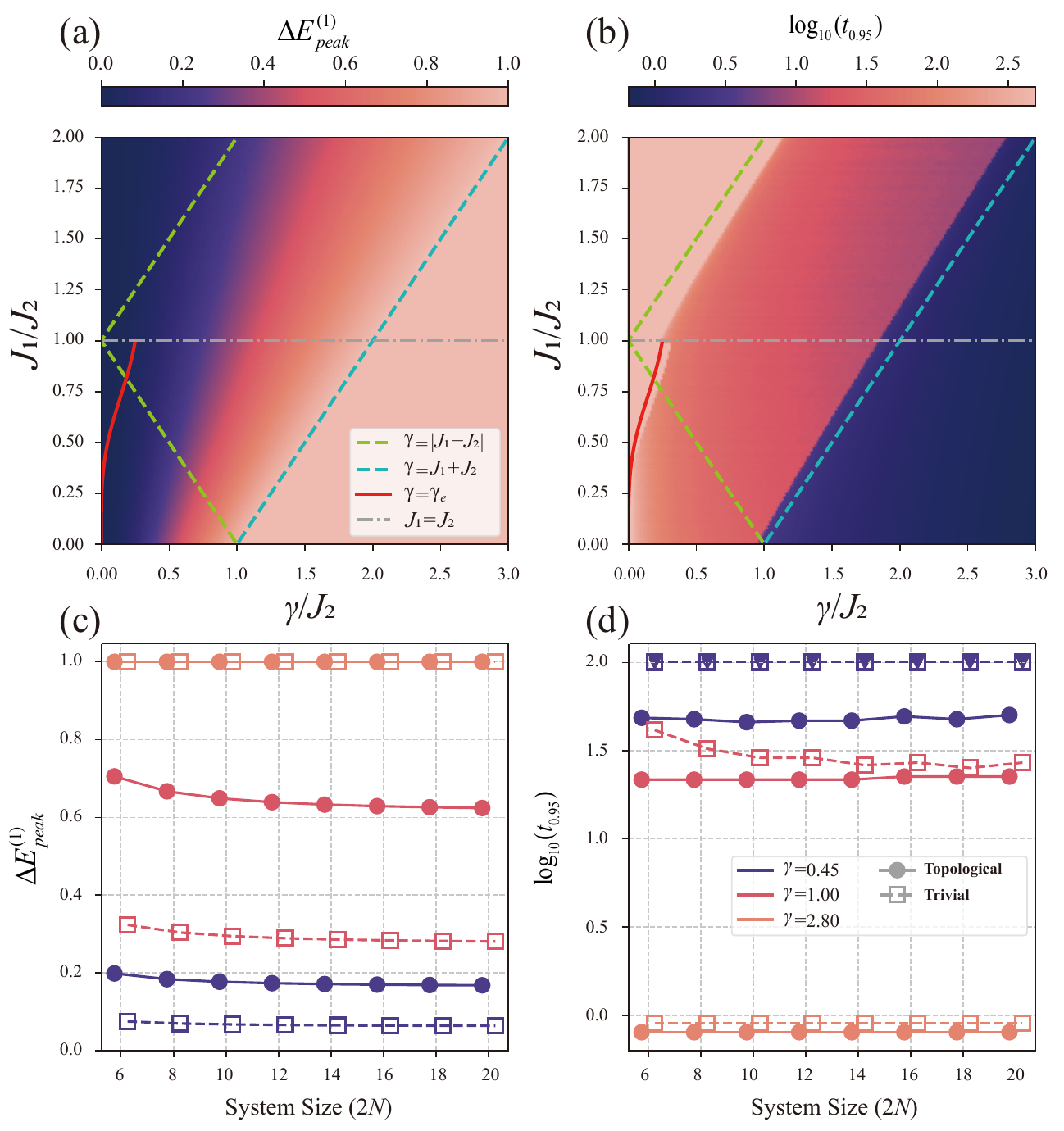}
    \caption{
    Dependence of the charging performance on system parameters of the $\mathcal{PT}$-symmetric SSH quantum battery.
    (a) Amplitude of the first energy peak $\Delta E_{\mathrm{peak}}^{(1)}$ in the $\gamma$--$J_{1}$ plane.
    (b) Logarithmic saturation time $\log_{10}(t_{0.95})$ in the same parameter space. The phase boundaries $J_{1}=J_{2}$ and the $\mathcal{PT}$-symmetry-breaking thresholds $\gamma_{e}$, $|J_{1}-J_{2}|$, and $J_{1}+J_{2}$ are indicated for reference.
    (c) System-size dependence of $\Delta E_{\mathrm{peak}}^{(1)}$ for three representative values of the gain-loss strength $\gamma/J_{2}=0.45$, $1.0$, and $2.8$.
    (d) Corresponding system-size dependence of the saturation time, shown as $\log_{10}(t_{0.95})$.
  In panels (c) and (d), solid curves correspond to the topological configuration ($J_{1}/J_{2}=0.5$) and dashed curves correspond to the trivial configuration ($J_{1}/J_{2}=1.5$).    
  }
    \label{fig:charging_metrics}
\end{figure}

\begin{figure*}[t]
    \centering
    \includegraphics[width=.92\linewidth]{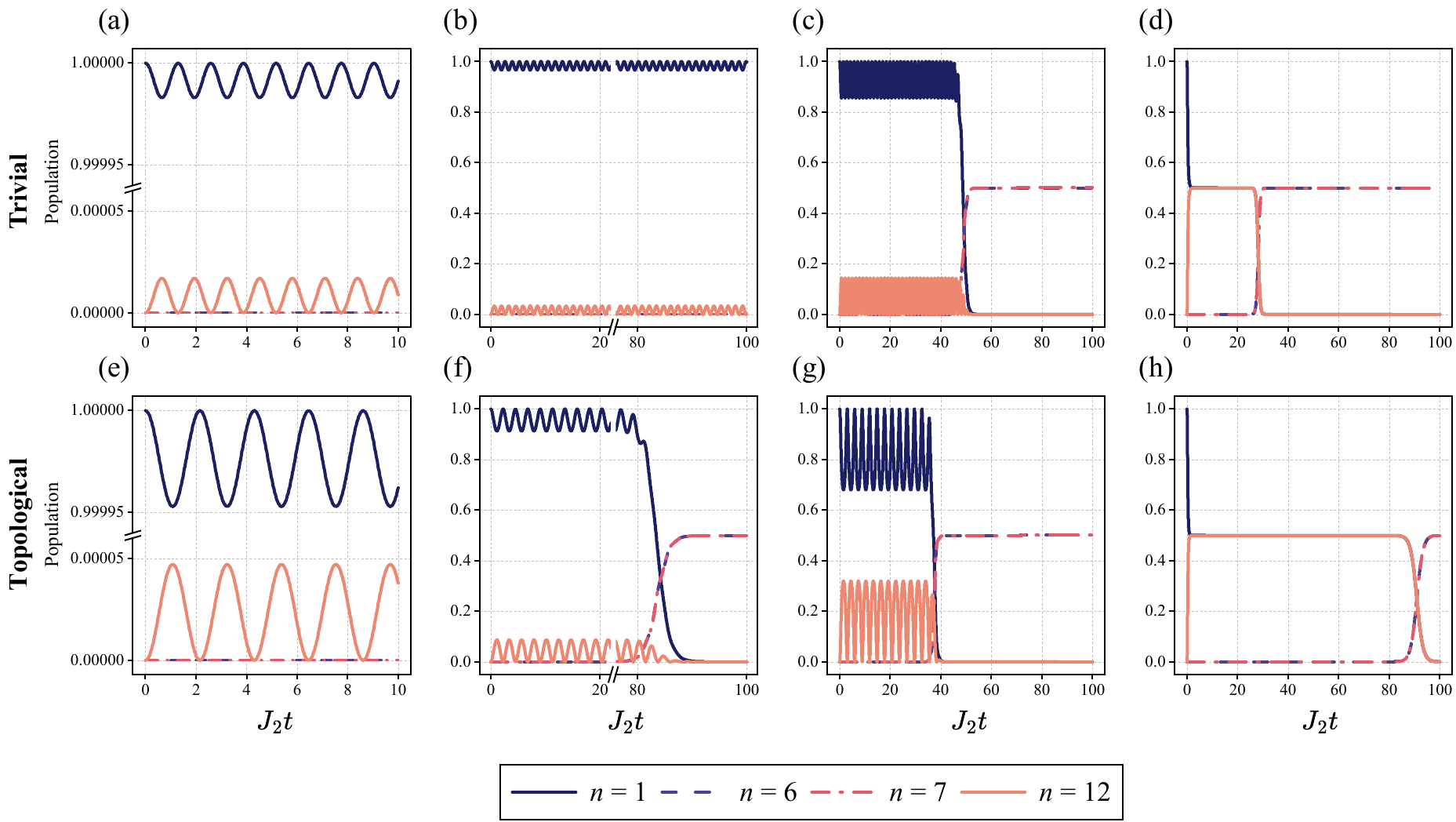}
    \caption{
    Time-dependent populations of selected eigenstates \(\phi_n\) of the SSH battery during \(\mathcal{PT}\)-symmetric charging.
    (a,e) Unbroken regime (\(\gamma/J_{2}=0.01\));
    (b,f) edge-state broken regime (\(\gamma/J_{2}=0.45\));
    (c,g) partially broken regime (\(\gamma/J_{2}=1.0\));
    (d,h) fully broken regime (\(\gamma/J_{2}=2.8\)).
    \rev{Top and bottom panels correspond to the trivial and topological phases, respectively.}
    \rev{Only the selected dominant populations are shown; the full population distribution generally includes additional states.}
    \rev{The observed transfer of weight among these selected states is consistent with the relative growth rates set by \(\operatorname{Im}E\).}
    }
    \label{fig:population_dynamics}
\end{figure*}

Together, $\Delta E_{\mathrm{peak}}^{(1)}$ and $t_{0.95}$ provide a comprehensive characterization of transient and steady-state behavior, enabling us to verify the generality of the charging trends observed in Sec.~\ref{sec:charging_dynamics} across the entire $\gamma$–$J_{1}$ parameter plane.
In fully oscillatory regimes where the battery never saturates, $t_{0.95}$ is effectively infinite, whereas in monotonic regimes $\Delta E_{\mathrm{peak}}^{(1)}$ is set to unity.
Since $t_{0.95}$ spans several orders of magnitude in the partially and fully broken regimes, we present $\log_{10}(t_{0.95})$ for clarity.

Figures~\ref{fig:charging_metrics}(a) and~\ref{fig:charging_metrics}(b) display these two metrics in the $\gamma$--$J_{1}$ plane.
Panel (a) shows the distribution of the first energy peak $\Delta E_{\mathrm{peak}}^{(1)}$, while panel (b) presents the logarithmic saturation time $\log_{10}(t_{0.95})$.
Both maps include the boundaries of the topological transition ($J_{1}=J_{2}$) and the $\mathcal{PT}$-symmetry-breaking thresholds $\gamma_{e}$, $|J_{1}-J_{2}|$, and $J_{1}+J_{2}$.
\rev{Throughout the parameter ranges shown in Figs.~\ref{fig:charging_metrics}(a) and \ref{fig:charging_metrics}(b), the topological phase exhibits larger \(\Delta E_{\mathrm{peak}}^{(1)}\) and shorter \(t_{0.95}\) than the trivial phase, indicating more favorable charging performance in these representative regimes.}

\rev{This difference is closely related to the topology-specific spectral structure introduced by the edge-state EP \(\gamma_{e}\), which appears at a much smaller gain-loss strength than the bulk thresholds.}
\rev{As a result, the topological configuration already displays distinct charging behavior in the edge-broken regime, where the edge-state pair is \(\mathcal{PT}\)-broken while the bulk modes remain real.}
\rev{The contrast between the regions below and above \(\gamma_{e}\) therefore highlights the influence of this additional topology-specific regime on the charging performance.}

\rev{Since the position of the edge-state EP depends exponentially on system size, we next investigate how the performance scales with the total number of sites \(2N\).}
\rev{Figures~\ref{fig:charging_metrics}(c) and \ref{fig:charging_metrics}(d) show the system-size dependence of \(\Delta E_{\mathrm{peak}}^{(1)}\) and \(\log_{10}(t_{0.95})\), respectively, for three representative non-Hermitian strengths \(\gamma/J_{2}=0.45\), \(1.0\), and \(2.8\).}
\rev{In all three cases, the topological and trivial configurations remain clearly distinguishable: the topological battery exhibits a higher first-peak energy and a shorter saturation time than the trivial one for the representative values of \(\gamma/J_{2}\) considered here.}
\rev{This shows that the topology-dependent difference in both transient and long-time charging performance is not a finite-size artifact, but persists as the system size increases.}

\rev{At \(\gamma/J_{2}=0.45\), the difference is particularly pronounced.}
\rev{The topological system lies in the edge-broken regime and achieves saturation, whereas the trivial system remains \(\mathcal{PT}\)-unbroken and fails to saturate, demonstrating a qualitative distinction in their dynamical behavior.}
\rev{As the system size increases, the critical value \(\gamma_{e}\) decreases, so that this topology-specific regime extends to weaker gain-loss strengths in the \(\gamma\)–\(J_{1}\) plane.}

\rev{Overall, these quantitative metrics confirm that the topological and trivial configurations exhibit systematically different charging characteristics, with the topological case showing more favorable transient and long-time performance in the representative regimes considered here.}

\subsection{Microscopic mechanism of the charging dynamics}
\label{sec:population_dynamics}

To uncover the microscopic origin of the charging behavior, we examine the time-dependent populations of the eigenstates of the SSH battery Hamiltonian during the evolution under the non-Hermitian charger.
We denote by $\phi_{1}$ and $\phi_{2N}$ the lowest- and highest-energy eigenstates of $H_{\mathrm{SSH}}$, respectively, and label the remaining eigenstates $\{\phi_{j}\}_{j=2}^{2N-1}$ in increasing energy order.
Because $\{\Gamma,H_{\mathrm{SSH}}\}=0$, the spectrum of $H_{\mathrm{SSH}}$ forms $\pm E_{j}$ pairs, and the drive term $i\gamma\Gamma$ couples each $+E_{j}$ state predominantly to its $-E_{j}$ partner.
This produces an effective two-level exchange in the weakly non-Hermitian regime.

The population dynamics across different values of $\gamma$ can be interpreted using the universal principles of non-Hermitian evolution summarized in Appendix~\ref{app:principles}.
Early-time behavior is controlled by the initial projections onto the eigenmodes, intermediate-time behavior by the gap between the leading imaginary parts, and long-time behavior by the mode with the largest imaginary part.

In the unbroken $\mathcal{PT}$ regime ($\gamma/J_{2}=0.01$), all eigenvalues are real and the chiral selection rule confines the dynamics essentially to the extremal pair $(\phi_{1},\phi_{2N})$.
Both phases therefore display small-amplitude, long-lived oscillations, as seen in Figs.~\ref{fig:population_dynamics}(a) and \ref{fig:population_dynamics}(e).
In a simple two-level approximation within this extremal subspace, the non-Hermitian drive $i\gamma\Gamma$ generates an off-diagonal coupling $\Gamma_{2N,1}=\langle\phi_{2N}|\Gamma|\phi_{1}\rangle$, and the maximal population transferred to $\phi_{2N}$ scales as $P_{2N}^{(\mathrm{max})}\propto \bigl(2\gamma|\Gamma_{2N,1}|/\Delta E\bigr)^{2}$, with $\Delta E=E_{2N}-E_{1}$.
Thus, the oscillation amplitude is set by the ratio $|\Gamma_{2N,1}|/|\Delta E|$, which can be confirmed numerically by comparing the topological and trivial configurations.
The same mechanism also underlies the systematically larger early-time oscillation amplitudes of the topological chain at higher gain–loss strengths, as seen by comparing Figs.~\ref{fig:population_dynamics}(f) and \ref{fig:population_dynamics}(g) with Figs.~\ref{fig:population_dynamics}(b) and \ref{fig:population_dynamics}(c).

At $\gamma/J_{2}=0.45$, the two phases separate due to the distinct fates of the edge doublet.
In the topological chain, the edge modes encounter the EP at the exponentially small threshold $\gamma_{e}$, causing one edge superposition to acquire a positive imaginary part and grow exponentially.
After normalization, this amplifying component rapidly dominates the population, leading to the takeover of the midgap states $(\phi_{6},\phi_{7})$ as shown in Fig.~\ref{fig:population_dynamics}(f).
The trivial chain lacks edge modes and remains unbroken at this $\gamma$, displaying only weak oscillations as in Fig.~\ref{fig:population_dynamics}(b).

When $\gamma/J_{2}=1.0$, both systems enter the bulk partially broken regime and possess a mixture of real and complex eigenvalues.
The short-time dynamics is still governed by oscillatory exchange among modes with real eigenvalues, whereas the long-time dynamics is dictated by the modes with the largest imaginary parts.
\rev{Because the topological chain already contains an edge-broken contribution in this parameter range, population is funneled into the amplifying subspace more efficiently than in the trivial chain.}
\rev{This difference in the dominant amplifying components is consistent with the faster saturation visible in Figs.~\ref{fig:population_dynamics}(c) and \ref{fig:population_dynamics}(g).}

\rev{In the fully broken regime (\(\gamma/J_{2}=2.8\)), all eigenvalues are complex and appear in conjugate pairs with opposite imaginary parts.}
\rev{Modes with positive imaginary parts are exponentially amplified, while their partners with negative imaginary parts are suppressed, and the normalized late-time dynamics is therefore governed by the former.}
\rev{For the parameters used in Fig.~\ref{fig:population_dynamics}, the largest imaginary parts are carried by modes near the band center, so both chains ultimately converge to the \((\phi_{6},\phi_{7})\) pair.}
\rev{Within each broken pair, the non-Hermitian drive combines coherent mixing with unequal amplification and decay, leading to a transient redistribution of population before the amplifying component prevails.}
\rev{Correspondingly, \(\Delta E(t)\) approaches its saturation value in both phases.}
\rev{The approach to this limit, however, still differs between the two configurations: in the trivial chain the gap between the two leading imaginary parts is larger, so the band-center pair overtakes competing modes more quickly, whereas in the topological chain several modes with comparable imaginary parts compete over a longer time window, delaying the emergence of the \((\phi_{6},\phi_{7})\) plateau.}
\rev{This delayed plateau does not imply slower charging in the topological case, because the stored energy has already reached a near-saturated value before the \((\phi_{6},\phi_{7})\) pair becomes dominant.}
\rev{The later crossover to the \((\phi_{6},\phi_{7})\) pair therefore reflects a long-time redistribution among amplifying eigenstates, rather than the onset of charging itself.}

In summary, the charging dynamics is governed by the interplay between chiral-selection-induced couplings, initial projections, and the hierarchy of imaginary parts of the eigenvalues.
\rev{These factors explain the qualitative differences between the trivial and topological configurations across the different non-Hermitian regimes considered here, while the edge-state EP provides a useful interpretation of the topology-specific dynamics in the edge-broken regime.}

\rev{\section{Lindblad validation of the topological enhancement}}
\label{sec:lindblad_validation}

\rev{
The $\mathcal{PT}$-symmetric non-Hermitian SSH Hamiltonian considered above is naturally understood as an effective description of an underlying open quantum system.
More specifically, it arises as the no-jump generator in a Lindblad quantum-trajectory picture.
This relation is important for the interpretation of the charging results obtained above: the non-Hermitian analysis captures the conditional amplification dynamics, while the full Lindblad equation determines whether the same topological advantage survives in the corresponding unconditional open-system evolution.

To make this connection explicit, we consider the Lindblad master equation
\begin{equation}
\dot{\rho}=-i[H_B,\rho]+\sum_\mu\left(L_\mu\rho L_\mu^\dagger-\frac12\{L_\mu^\dagger L_\mu,\rho\}\right),
\label{eq:lindblad_master}
\end{equation}
where \(H_B\) is the SSH battery Hamiltonian and the jump operators encode gain and loss on the two sublattices.
For the full-sublattice \(\mathcal{PT}\)-symmetric setting studied here, we choose
\begin{equation}
L_{a,n}=\sqrt{\kappa}\,a_n^\dagger,\qquad
L_{b,n}=\sqrt{\kappa}\,b_n,
\label{eq:jump_ops_main}
\end{equation}
namely, gain on all \(A\)-sublattice sites and loss on all \(B\)-sublattice sites \cite{dast_quantum_2014}.
In the quantum-trajectory formulation, the corresponding no-jump evolution is generated by \cite{nakanishi_pt_2022, he_quantum_2015, huber_emergence_2020}
\begin{equation}
H_{\rm eff}=H_B-\frac{i}{2}\sum_\mu L_\mu^\dagger L_\mu.
\label{eq:Heff_from_lindblad}
\end{equation}
Substituting the above jump operators into this expression yields the \(\mathcal{PT}\)-symmetric Su-Schrieffer-Heeger (PTSSH) Hamiltonian used in Eq.~(\ref{eq:H_PT}), up to a term proportional to the identity.
Upon identifying \(\gamma=\kappa/2\), the effective non-Hermitian model is therefore recovered as the conditional no-jump description associated with the same gain/loss processes.
The identity term only changes the overall no-jump norm and does not affect the normalized conditional dynamics or the spectral structure.

This correspondence clarifies the role of the non-Hermitian description.
The PTSSH Hamiltonian is not introduced independently of the open-system setting; rather, it describes the conditional dynamics in which no quantum jump occurs.
The full Lindblad equation, by contrast, governs the unconditional evolution obtained after averaging over all jump trajectories.
The question addressed in this section is therefore not whether the Lindblad and non-Hermitian descriptions are identical, but whether the topological charging advantage identified in the conditional dynamics remains visible once the jump processes are restored.

To keep the comparison as close as possible to the previous analysis, we work in the same single-excitation setting.
Concretely, we truncate the Hilbert space to the vacuum state together with the full single-excitation sector.
Within this space, the loss operator \(b_n\) maps a single excitation to the vacuum, while the gain operator \(a_n^\dagger\) injects an excitation from the vacuum into the chain.
This provides the minimal open-system extension of the PTSSH battery and allows a direct comparison between the conditional and unconditional pictures.

Under unconditional Lindblad evolution, the battery state is generally mixed.
In this situation, the mean energy alone is not sufficient to quantify battery performance, because part of the stored energy may already be passive and hence unavailable for work extraction.
We therefore distinguish between the total stored energy and the extractable part of that energy.

The stored energy is characterized by the energy above the ground state,
\begin{equation}
\Delta E(t)=E(t)-E_{\min},\qquad
E(t)=\mathrm{Tr}[H_B\rho(t)],
\label{eq:available_energy_def}
\end{equation}
where \(E_{\min}\) is the ground-state energy of \(H_B\).
This quantity measures how much energy has been accumulated in the battery relative to its lowest-energy configuration.
In the present mixed-state setting, this ground-state reference is more natural than the initial-state reference used for the conditional pure-state dynamics above, because the long-time Lindblad steady state is interpreted as an energy-storage state rather than as a coherent trajectory relative to a fixed initial vector.

To quantify the useful part of the stored energy, we compute the ergotropy
\begin{equation}
W(t)=\mathrm{Tr}[H_B\rho(t)]-\mathrm{Tr}[H_B\rho_{\rm pas}(t)],
\label{eq:ergotropy_def_main}
\end{equation}
where \(\rho_{\rm pas}(t)\) is the passive state associated with \(\rho(t)\).
Physically, \(\rho_{\rm pas}\) is obtained by rearranging the eigenvalues of \(\rho\) onto the eigenstates of \(H_B\) in increasing-energy order, thereby minimizing the energy within the unitary orbit of \(\rho\).
The ergotropy thus measures the non-passive part of the stored energy, i.e., the part that can actually be extracted as useful work by cyclic unitary operations.
By construction,
\begin{equation}
0\le W(t)\le \Delta E(t).
\label{eq:ergotropy_bound}
\end{equation}
This inequality is useful for interpreting the numerical results below, because it directly separates the stored energy into extractable and passive contributions.

A convenient dimensionless measure is the extractable fraction
\begin{equation}
\eta_W(t)=\frac{W(t)}{\Delta E(t)},
\label{eq:eta_def_main}
\end{equation}
which characterizes the quality of charging.
When \(\eta_W\) approaches unity, almost all stored energy is available for work extraction; when \(\eta_W\) is small, a substantial part of the stored energy is passive.
In the present context, \(\eta_W\) is particularly informative because it reveals whether topology improves not only the amount of energy stored, but also the fraction of that energy that remains useful under open-system dynamics.

In the following, we focus on the long-time limit of the Lindblad dynamics.
The corresponding steady-state quantities are defined as
\begin{equation}
\begin{aligned}
\Delta E_{\rm ss} &= \Delta E(t\to\infty), \\
W_{\rm ss} &= W(t\to\infty), \\
\eta_{W,{\rm ss}} &= \eta_W(t\to\infty).
\end{aligned}
\label{eq:steady_state_quantities}
\end{equation}

For the comparison in this section, we evaluate these quantities using the same normalized battery Hamiltonian convention adopted above.
This keeps the Lindblad results on the same energy scale as the conditional non-Hermitian results and removes the trivial influence of bandwidth differences between the topological and trivial parameter sets.
The corresponding steady-state results are shown in Fig.~\ref{fig:norm_energy_work_eta}.
Figure \ref{fig:norm_energy_work_eta}(a) displays both the stored energy \(\Delta E_{\rm ss}\) and the steady-state ergotropy \(W_{\rm ss}\), while Fig.~\ref{fig:norm_energy_work_eta}(b) shows the extractable fraction \(\eta_{W,{\rm ss}}\), all as functions of the gain/loss strength \(\gamma\).

\begin{figure}[t]
    \centering
    \includegraphics[width=\linewidth]{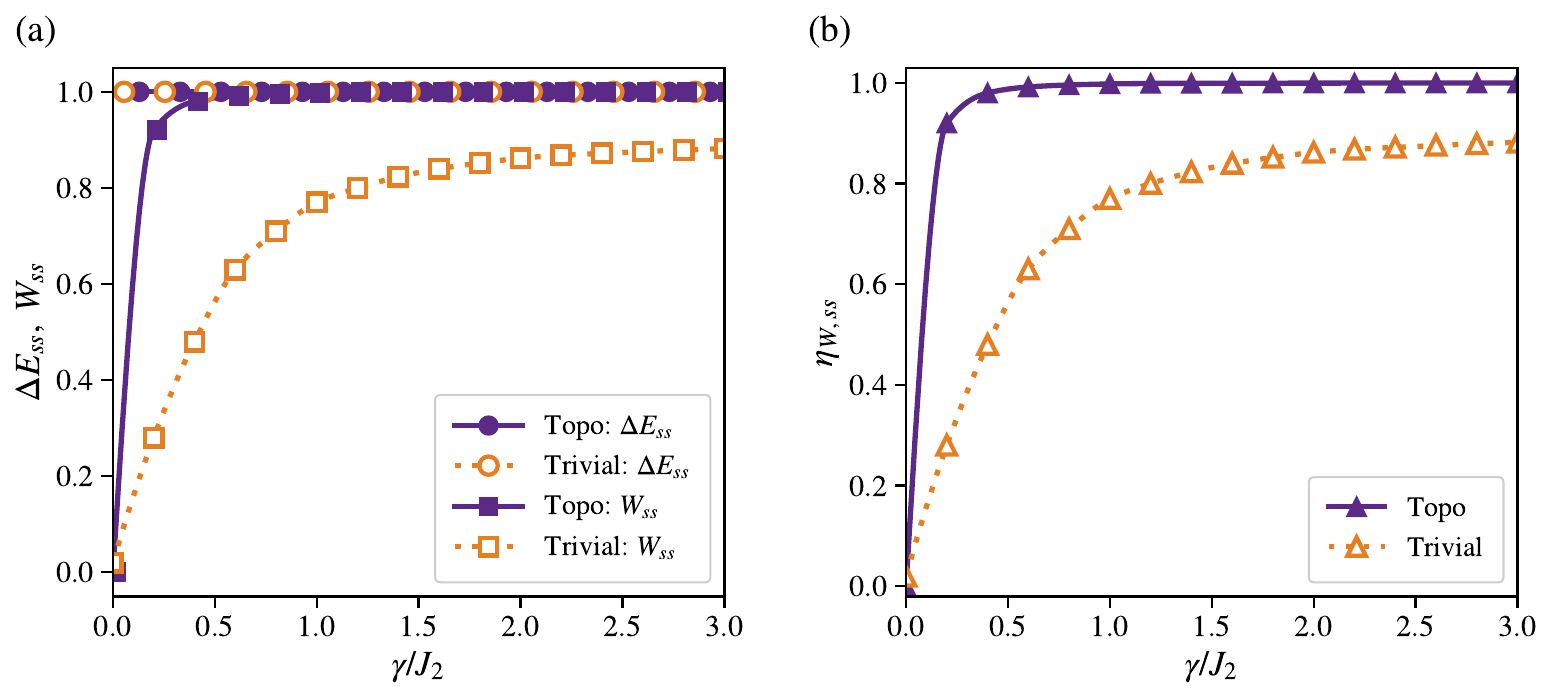}
    \caption{
    Steady-state battery performance under the unconditional Lindblad dynamics in the single-excitation sector, shown as functions of the gain--loss strength $\gamma$.
    (a) Normalized steady-state stored energy $\Delta E_{ss}$ and ergotropy $W_{ss}$.
    (b) Normalized steady-state extractable fraction $\eta_{W,ss}=W_{ss}/\Delta E_{ss}$.
    The topological and trivial configurations are shown by solid filled purple curves and dotted open orange curves, respectively.
    In panel (a), circles and squares denote $\Delta E_{ss}$ and $W_{ss}$, respectively.
    In panel (b), triangles denote $\eta_{W,ss}$.
    }
    \label{fig:norm_energy_work_eta}
\end{figure}

Several features are immediately visible.
First, both parameter sets acquire a finite steady-state energy above the ground-state reference under the unconditional Lindblad dynamics.
Second, the topological parameter set yields a larger steady-state ergotropy over a broad range of \(\gamma\), even when the stored energy itself is comparable to that of the trivial phase.
This shows that the topological advantage is not merely an increase in the total amount of stored energy.
Rather, topology favors the accumulation of energy in a less passive and hence more useful form.

This interpretation is reinforced by the extractable fraction shown in Fig.~\ref{fig:norm_energy_work_eta}(b).
The topological phase consistently exhibits a larger \(\eta_{W,{\rm ss}}\), indicating that a larger portion of the stored energy remains available for work extraction in the steady state.
From the viewpoint of open-system battery performance, this is the central message of the Lindblad validation: the topological advantage survives the inclusion of quantum jumps and remains clearly visible at the level of extractable work.

A concise physical interpretation can be given as follows.
The conditional non-Hermitian dynamics discussed above already shows that the topological parameter set favors a more selective amplification pattern associated with the PTSSH structure.
Once the quantum jumps are restored, this conditional mechanism is no longer preserved trajectory by trajectory, because the unconditional Lindblad evolution mixes different jump histories and generally drives the battery state into a mixed steady state.
However, our results also show that the topological regime is less prone to quantum jumps than the trivial one, in the sense that its no-jump survival probability remains significantly larger over the relevant parameter range.
This makes the topological dynamics more robust against the degradation induced by jumps, so that the useful part of the conditional charging advantage can still be retained after averaging over trajectories.
As a result, the topological regime continues to accumulate a larger fraction of its stored energy in a non-passive form, which is why its advantage remains most visible in the ergotropy and in the extractable fraction, rather than in the stored energy alone.

The physical picture that emerges is therefore the following.
The conditional non-Hermitian dynamics reveals the amplification mechanism and its topological dependence, while the unconditional Lindblad dynamics confirms that this mechanism is not washed out by the jump processes.
Instead, topology continues to enhance the work-like component of the stored energy.
In this sense, the Lindblad analysis provides an open-system validation of the non-Hermitian charging picture developed above.

For completeness, the corresponding results in the unnormalized energy scale are presented in Appendix~\ref{app:unnormalized_lindblad}.
Those results make explicit the absolute energy and work scales, and show how bandwidth differences influence the absolute values of the stored energy and ergotropy.
The normalized results discussed here are the appropriate ones for isolating the intrinsic topological enhancement independently of such scale effects.
}

~
\section{Conclusions}

\rev{In this work, we have studied the \(\mathcal{PT}\)-symmetric Su-Schrieffer-Heeger (PTSSH) model as a prototype of a non-Hermitian quantum battery.}  
\rev{The coexistence of topology and balanced gain-loss gives rise to both bulk and edge exceptional points (EPs), which organize the charging dynamics across different non-Hermitian regimes.}  

\rev{An edge-state EP, present only in the topological phase, appears at a smaller gain-loss strength than the bulk thresholds and gives rise to an additional edge-broken regime absent in the trivial case.}  
\rev{This topology-specific spectral structure contributes to the distinct charging behavior and enhanced performance of the topological configuration.}

\rev{Quantitative analysis of the stored energy \(\Delta E(t)\) and its derived metrics, the first-peak amplitude and the saturation time, confirms that the topological configuration consistently achieves higher transient energy and quicker saturation in the representative non-Hermitian regimes considered here.}  
\rev{Microscopic population dynamics show that, once the edge mode is broken, the eigenstate with the largest imaginary part of the eigenvalue eventually dominates the evolution, providing a microscopic picture of how energy accumulation is favored in the topological configuration.}

\rev{Further analysis based on the Lindblad master equation reveals that the corresponding unconditional open-system dynamics remains consistent with the effective non-Hermitian no-jump description, while also clarifying how the topological edge modes influence the steady-state energy and extractable work.}
\rev{These findings reinforce the robustness of the topological advantages and demonstrate how Lindblad dynamics modify the energy extraction efficiency in the presence of dissipation, further supporting the applicability of \(\mathcal{PT}\)-symmetric quantum batteries for practical use.}

\rev{These results demonstrate that topology and non-Hermiticity can act cooperatively to enhance quantum energy storage, offering a viable route toward designing high-performance quantum batteries based on non-Hermitian control.}  
\rev{Future extensions may explore interacting or many-body realizations, where collective charging and dissipation could yield new scaling behaviors.}  
\rev{Possible experimental implementations in photonic lattices, superconducting circuits, or cold-atom arrays would enable direct tests of the predicted edge-induced enhancement and probe the robustness of topological advantages in non-Hermitian quantum devices \cite{Mei_Robust_2018, Zeuner_Observation_2015, Lin_Dynamic_2021}.}

\begin{acknowledgments}
This work was supported by the National Natural Science Foundation of China (Grant No. 12205383, 12274470) and the Hunan Provincial Natural Science Foundation (Grant No. 2024JJ6483). 
A.-L. Zhou was also supported by the Fundamental Research Funds for the Central Universities of Central South University under Grant No.1053320214335. 
\end{acknowledgments}

\appendix

\section{Chiral symmetry and transition matrix elements}
\label{app:chiral}

In this appendix, we clarify how the chiral symmetry of the Hermitian SSH Hamiltonian constrains its spectrum and the matrix elements of the sublattice operator $\Gamma$, and how these constraints underlie the selection rules used in the main text.

We recall that the SSH Hamiltonian $H_{\mathrm{SSH}}$ introduced in Eq.~(\ref{eq:H_SSH}) acts on a chain of $N$ unit cells with sublattice operators $a_{n}$ and $b_{n}$.
The sublattice (chiral) operator is defined as
\begin{equation}
\Gamma
  = \sum_{n=1}^{N}
      (a_{n}^{\dagger}a_{n}-b_{n}^{\dagger}b_{n}) .
\end{equation}
A straightforward calculation shows that
\begin{equation}
\{\Gamma,H_{\mathrm{SSH}}\}
  = \Gamma H_{\mathrm{SSH}} + H_{\mathrm{SSH}}\Gamma
  = 0,
\end{equation}
so $H_{\mathrm{SSH}}$ possesses an exact chiral symmetry.

Let $|\phi_{j}\rangle$ be the normalized eigenstates of $H_{\mathrm{SSH}}$, ordered so that $H_{\mathrm{SSH}}|\phi_{j}\rangle = E_{j}|\phi_{j}\rangle$ with $E_{1}\leq E_{2}\leq\cdots\leq E_{2N}$.
Using $\{\Gamma,H_{\mathrm{SSH}}\}=0$, we obtain
\begin{align}
0
 &= \langle\phi_{m}|\{\Gamma,H_{\mathrm{SSH}}\}|\phi_{n}\rangle \nonumber\\
 &= (E_{m}+E_{n})\,\langle\phi_{m}|\Gamma|\phi_{n}\rangle .
\end{align}
For nondegenerate eigenvalues, this implies the ``chiral selection rule''
\begin{equation}
(E_{m}+E_{n})\neq 0
\quad\Rightarrow\quad
\langle\phi_{m}|\Gamma|\phi_{n}\rangle = 0 .
\end{equation}
In other words, the sublattice operator $\Gamma$ only couples eigenstates whose energies are opposite, $E_{m}=-E_{n}$.
This condition is a direct consequence of chiral symmetry and does not depend on microscopic details.

Because the SSH spectrum is symmetric around zero energy, we can pair the eigenstates as $(|\phi_{j}\rangle,|\phi_{\bar{j}}\rangle)$ with $E_{\bar{j}}=-E_{j}$.
For an open chain of $2N$ sites it is convenient to choose the ordering such that $\bar{j}=2N+1-j$, so that $(|\phi_{1}\rangle,|\phi_{2N}\rangle)$, $(|\phi_{2}\rangle,|\phi_{2N-1}\rangle)$, and so on form chiral partner pairs.
Within this convention, the matrix elements of $\Gamma$ in the energy eigenbasis,
\begin{equation}
M_{mn} = \langle\phi_{m}|\Gamma|\phi_{n}\rangle ,
\end{equation}
are strongly concentrated near the anti-diagonal $m=\bar{n}$, while elements with $m\neq\bar{n}$ are suppressed by the selection rule.

For the finite-size parameters used in the main text, we have verified numerically that the largest matrix element is
\begin{equation}
|M_{1,2N}| = |\langle\phi_{1}|\Gamma|\phi_{2N}\rangle|
 \simeq 1 ,
\end{equation}
whereas $M_{1,2}$, $M_{1,3}$, and other off-pair terms are typically at least an order of magnitude smaller.
Similar behavior is found for other chiral pairs $(\phi_{j},\phi_{\bar{j}})$.
This pattern implies that, in the weakly non-Hermitian regime where the drive term $i\gamma\Gamma$ acts as a perturbation, the dominant transition channel induced by $\Gamma$ connects the lowest- and highest-energy eigenstates, while additional couplings within the same energy sign are strongly suppressed.

In the presence of the non-Hermitian gain–loss term, the full $\mathcal{PT}$-symmetric Hamiltonian
\begin{equation}
H_{\mathcal{PT}}
 = H_{\mathrm{SSH}} + i\gamma\,\Gamma
\end{equation}
no longer anticommutes with $\Gamma$.
Indeed,
\begin{equation}
\{\Gamma,H_{\mathcal{PT}}\}
 = 2i\gamma\,\mathbb{I} \neq 0 .
\end{equation}
The chiral symmetry is thus explicitly broken once $\gamma\neq0$.
Nevertheless, the matrix structure of $\Gamma$ in the $\{|\phi_{j}\rangle\}$ basis retains its pronounced anti-diagonal character, and the extremal pair $(\phi_{1},\phi_{2N})$ continues to play a special role.
This \emph{spectral memory} explains why, even when non-Hermiticity is present, the early-time charging dynamics is dominated by transitions between the lowest- and highest-energy eigenstates, as discussed in Sec.~\ref{sec:population_dynamics}.

\section{Equivalence between ergotropy and charging energy}
\label{app:ergotropy}

For completeness, we provide here a derivation showing that the ergotropy is exactly equal to the stored energy for the charging protocol considered in this work.
The ergotropy \cite{alicki_entanglement_2013, allahverdyan_maximal_2004, Yang_Battery_2023} of a quantum battery is defined as
\begin{equation}
\mathcal{W}(t)
 = \mathrm{Tr}[H_B\rho(t)]
   - \min_{U}\,\mathrm{Tr}\!\big[H_B\,U\rho(t)U^{\dagger}\big],
\label{eq:ergotropy_def}
\end{equation}
where $H_B$ denotes the battery Hamiltonian and $\rho(t)$ is the density matrix of the battery at time $t$.
The minimum in Eq.~(\ref{eq:ergotropy_def}) is achieved for the so-called \emph{passive state},
\begin{equation}
\rho_{\mathrm{p}}
 = \sum_{k} r_{k}^{\downarrow}
   \,|\varepsilon_{k}^{\uparrow}\rangle
    \langle \varepsilon_{k}^{\uparrow}| ,
\end{equation}
where $r_{k}^{\downarrow}$ are the eigenvalues of $\rho(t)$ arranged in nonincreasing order and $|\varepsilon_{k}^{\uparrow}\rangle$ are the eigenstates of $H_B$ ordered by increasing eigenvalues $E_{k}$.

In our model, the battery Hamiltonian coincides with the Hermitian SSH Hamiltonian $H_{\mathrm{SSH}}$ introduced in Sec.~\ref{sec:SSHmodel}, and its eigenstates are denoted by $\{|\phi_{j}\rangle\}$ with eigenvalues $\{E_{j}\}$ ordered such that $E_{1}\leq \cdots \leq E_{2N}$.
The ground state of the battery is therefore $|G\rangle=|\phi_{1}\rangle$, and the initial state is prepared as the corresponding projector,
\begin{equation}
\rho(0)
 = |G\rangle\langle G|
 = |\phi_{1}\rangle\langle\phi_{1}| ,
\end{equation}
which represents a pure state with minimum energy
\begin{equation}
E_B(0)
 = \langle G|H_B|G\rangle
 = E_{1} .
\end{equation}

During the non-Hermitian charging process, the battery evolves according to the normalized effective dynamics defined in Eq.~(\ref{eq:normalized_rho}).
Since this map is generated by $H_{\mathcal{PT}}$ acting on a pure initial state and involves a renormalization at each time step, the density matrix remains pure for all $t$,
\begin{equation}
\rho(t)
 = |\psi(t)\rangle\langle\psi(t)| .
\end{equation}
Hence, the spectrum of $\rho(t)$ contains a single nonzero eigenvalue equal to one, independently of the detailed superposition $|\psi(t)\rangle$.

Because the passive state is obtained by rearranging the eigenvalues of $\rho(t)$ in a basis that minimizes the energy, and there is only one nonzero eigenvalue, the passive counterpart of $\rho(t)$ is simply the ground-state projector,
\begin{equation}
\rho_{\mathrm{p}}
 = |G\rangle\langle G|
 = |\phi_{1}\rangle\langle\phi_{1}| .
\end{equation}
Substituting this $\rho_{\mathrm{p}}$ into Eq.~(\ref{eq:ergotropy_def}) yields
\begin{align}
\mathcal{W}(t)
 &= \mathrm{Tr}[H_B\rho(t)]
    - \mathrm{Tr}[H_B|G\rangle\langle G|]
 \notag\\
 &= E_B(t) - E_B(0)
 \notag\\
 &\equiv \Delta E(t) ,
\end{align}
where $E_B(t)=\mathrm{Tr}[H_B\rho(t)]$ is the instantaneous energy of the battery.
Therefore, for a pure initial state undergoing the normalized non-Hermitian evolution of Eq.~(\ref{eq:normalized_rho}) under $H_{\mathcal{PT}}$, the ergotropy is exactly equal to the charging energy,
\begin{equation}
\mathcal{W}(t)
 = \Delta E(t) .
\end{equation}
This result implies that, within the no-jump effective description adopted in the main text, the stored energy $\Delta E(t)$ alone faithfully characterizes the extractable work of the $\mathcal{PT}$-symmetric SSH quantum battery.

\section{Universal principles of non-Hermitian dynamics}
\label{app:principles}

In this appendix, we summarize three general principles that govern the time evolution of diagonalizable non-Hermitian systems.
These principles follow directly from the spectral decomposition of the evolution operator and are independent of the microscopic details of the Hamiltonian.

We consider a non-Hermitian Hamiltonian $H$ with right eigenvectors $|\phi_{j}\rangle$ and complex eigenvalues $E_{j}=\omega_{j}+i\lambda_{j}$, where $\omega_{j}$ and $\lambda_{j}$ denote the real and imaginary parts, respectively.
We assume that $H$ is diagonalizable, i.e., that the system is not tuned exactly to an exceptional point, so that the set $\{|\phi_{j}\rangle\}$ spans the Hilbert space.
Starting from an initial state $|\psi(0)\rangle=\sum_{j}a_{j}|\phi_{j}\rangle$, the unnormalized state evolves as
\begin{equation}
|\psi_{\mathrm{un}}(t)\rangle
 = e^{-iHt}|\psi(0)\rangle
 = \sum_{j}a_{j}\,e^{-iE_{j}t}\,|\phi_{j}\rangle .
\end{equation}
In many applications, including the main text, one works with the normalized state
\begin{equation}
|\psi(t)\rangle
 = \frac{|\psi_{\mathrm{un}}(t)\rangle}
        {\| |\psi_{\mathrm{un}}(t)\rangle \|} ,
\end{equation}
or with the corresponding density matrix $\rho(t)=|\psi(t)\rangle\langle\psi(t)|$.

The \emph{long-time behavior} is governed solely by the imaginary parts $\{\lambda_{j}\}$.
Let $\lambda_{\max}=\max_{j}\lambda_{j}$ be the largest growth rate and assume it is nondegenerate.
Then, for $t\to\infty$, the contribution of the eigenmode with $\lambda_{\max}$ dominates $|\psi_{\mathrm{un}}(t)\rangle$, and after normalization the state converges to the corresponding eigenvector,
\begin{equation}
|\psi(t)\rangle
 \xrightarrow[t\to\infty]{}
 |\phi_{\max}\rangle ,
\end{equation}
where $|\phi_{\max}\rangle$ is the eigenstate with $\operatorname{Im}E_{\max}=\lambda_{\max}$.
If several modes share the same maximal imaginary part, the long-time state resides in the subspace spanned by these modes, but the final superposition is determined by the initial coefficients $\{a_{j}\}$.

The \emph{relaxation timescale} is set by the gap between the leading growth rates.
Denoting by $\lambda_{2}$ the second-largest imaginary part, the difference
\begin{equation}
\Delta\lambda
 = \lambda_{\max}-\lambda_{2}
\end{equation}
controls how quickly the dominant mode overwhelms its competitors.
A larger gap $\Delta\lambda$ leads to rapid convergence to the asymptotic state, while nearly degenerate growth rates produce a prolonged competition among several modes.
This observation underlies the estimate
\begin{equation}
\tau^{-1}
 \sim \lambda_{\max}-\lambda_{2} ,
\end{equation}
used in the main text to relate spectral properties to charging times.

At \emph{early times}, however, the dynamics is controlled primarily by the initial projections $a_{j}$ rather than by the growth rates.
Modes with large initial overlap can dominate the short-time evolution even if they do not possess the largest imaginary parts.
As time increases, the exponential factors $e^{\lambda_{j}t}$ eventually amplify the contribution of the leading mode(s), and the role of the initial coefficients becomes subdominant.

These three ingredients---initial overlaps, the gap between the leading imaginary parts, and the identity of the mode(s) with maximal $\lambda_{j}$---provide a universal framework for analyzing non-Hermitian dynamics.
They explain in a model-independent way the transient and asymptotic features of the population dynamics discussed in Sec.~\ref{sec:population_dynamics}, including the dominance of the extremal SSH eigenpair at weak non-Hermiticity, \rev{the topology-dependent saturation behavior associated with the edge-broken regime in the topological phase}, and the late-time selection of band-center modes in the fully broken regime.

\section{\rev{Unnormalized energy scale and practical output in Lindblad dynamics}}
\label{app:unnormalized_lindblad}

\rev{
The normalized comparison in the main text is intended to isolate the intrinsic difference between the topological and trivial parameter sets by removing the trivial effect of different bandwidths.
For completeness, it is also useful to examine the same steady-state quantities in the unnormalized energy scale, namely with the battery Hamiltonian taken directly as \(H_B=H_{\rm SSH}\).
This representation restores the physical energy scale set by the couplings \(J_1\) and \(J_2\), and therefore provides a more direct estimate of the absolute output of the battery.

\begin{figure}[t]
    \centering
    \includegraphics[width=\linewidth]{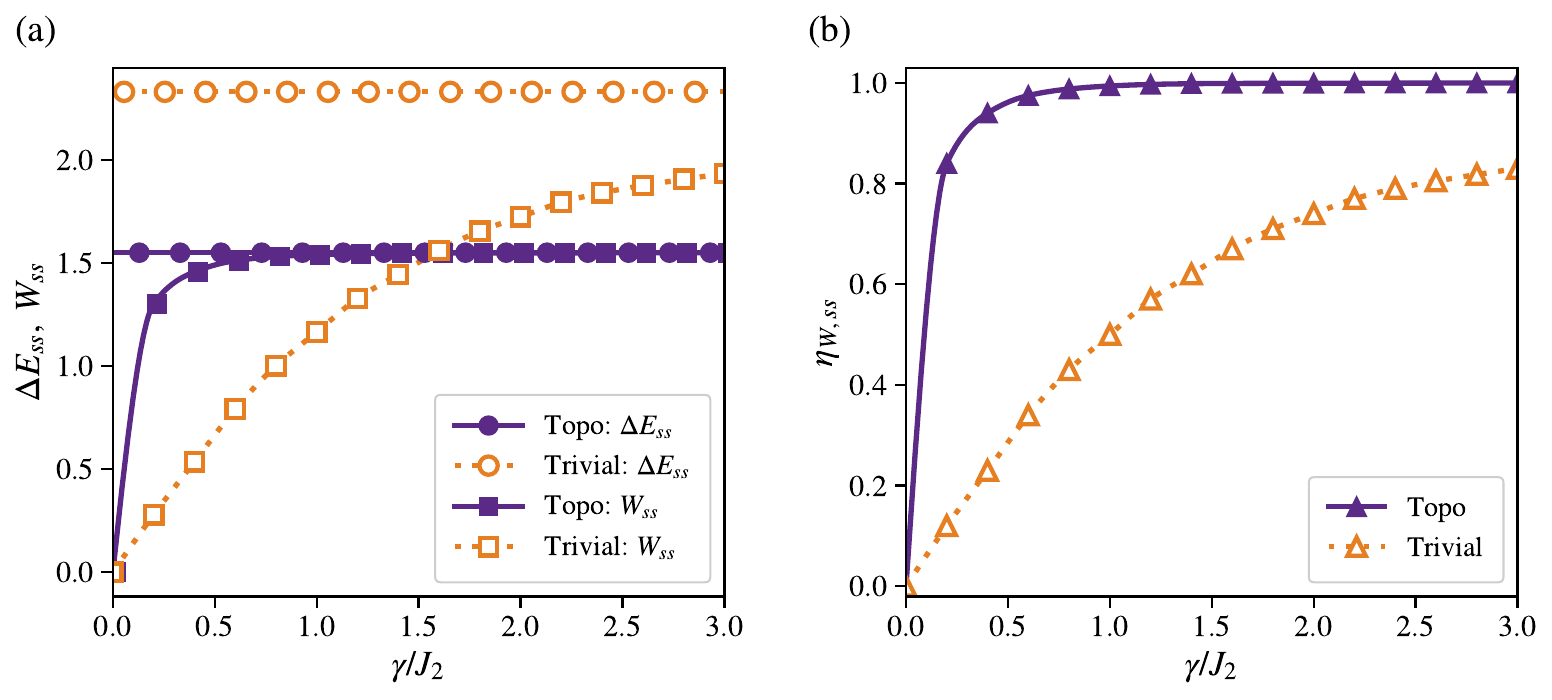}
    \caption{
    Unnormalized steady-state battery performance under the unconditional Lindblad dynamics in the single-excitation sector, shown as functions of the gain--loss strength $\gamma$.
    (a) Unnormalized steady-state stored energy $\Delta E_{ss}$ and ergotropy $W_{ss}$.
    (b) Unnormalized steady-state extractable fraction $\eta_{W,ss}=W_{ss}/\Delta E_{ss}$.
    The topological and trivial configurations are shown by solid filled purple curves and dotted open orange curves, respectively.
    In panel (a), circles and squares denote $\Delta E_{ss}$ and $W_{ss}$, respectively.
    In panel (b), triangles denote $\eta_{W,ss}$.
    }
    \label{fig:unnorm_energy_work_eta}
\end{figure}

The corresponding results are shown in Fig.~\ref{fig:unnorm_energy_work_eta}.
As in the normalized case, Fig.~\ref{fig:unnorm_energy_work_eta}(a) displays both the stored energy above the ground state, \(\Delta E_{\rm ss}=E_{\rm ss}-E_{\min}\), and the steady-state ergotropy \(W_{\rm ss}\), while Fig.~\ref{fig:unnorm_energy_work_eta}(b) shows the extractable fraction \(\eta_{W,{\rm ss}}=W_{\rm ss}/\Delta E_{\rm ss}\).
In this representation, the left panel characterizes the absolute output of the battery, whereas the right panel characterizes the fraction of that output that remains available for work extraction.

The unnormalized comparison contains two effects at the same time.
On the one hand, it still reflects the dynamical distinction between the topological and trivial regimes, namely how effectively the Lindblad evolution stores energy in a non-passive form.
On the other hand, it also contains the difference in the intrinsic energy scales of the corresponding SSH Hamiltonians, since the two parameter sets generally have different bandwidths.
As a result, the absolute values of \(\Delta E_{\rm ss}\) and \(W_{\rm ss}\) should not be interpreted as purely dynamical indicators: they also depend on the overall energy capacity set by the Hamiltonian itself.

This point is directly visible in the Fig.~\ref{fig:unnorm_energy_work_eta}(a).
In the topological regime, the ergotropy remains close to the stored energy over a broad range of \(\gamma\), which means that most of the accumulated energy is retained in a work-extractable form.
In the trivial regime, by contrast, the absolute stored energy can become larger in some parameter ranges, but the separation between \(W_{\rm ss}\) and \(\Delta E_{\rm ss}\) is also larger, indicating that a greater portion of the injected energy is converted into passive energy.
The unnormalized plot therefore shows that a larger absolute energy scale does not necessarily imply a better battery from the viewpoint of useful work extraction.

Figure \ref{fig:unnorm_energy_work_eta}(b) makes this distinction particularly clear.
Even when the absolute stored energy differs significantly between the two parameter sets, the extractable fraction \(\eta_{W,{\rm ss}}\) still reveals how efficiently that stored energy is organized into a work-like form.
From this perspective, the topological regime remains advantageous over a broad range of \(\gamma\), because a larger fraction of the injected energy stays extractable rather than becoming passive.
This is consistent with the greater robustness of the topological regime against jump-induced degradation discussed in the main text.

The role of the present unnormalized comparison is therefore complementary to that of the normalized results in the main text.
The normalized plots are the appropriate ones for identifying the intrinsic topological enhancement independently of bandwidth differences.
The unnormalized plots, by contrast, show how that enhancement appears once the physical energy scales of the two Hamiltonians are restored.
Taken together, the two representations lead to a consistent picture: topology improves the quality of stored energy in a robust way, while the absolute amount of stored energy and extractable work can additionally depend on the underlying bandwidth of the battery Hamiltonian.
}

\end{document}